\documentclass[final,5p,times,twocolumn,12pt]{elsarticle}

\usepackage{amsmath}
\usepackage{epstopdf}
\usepackage{graphicx}
\usepackage{bm,color,bbm}
\usepackage{hyperref, mathtools}
\usepackage{caption}
\usepackage{subcaption}

\newcommand{\beq}{\begin{equation}}
\newcommand{\eeq}{\end{equation}}
\newcommand{\bqa}{\begin{eqnarray}}
\newcommand{\eqa}{\end{eqnarray}}

\journal{Physics Letters A}

\begin{document}

\begin{frontmatter}



\title{Improving Einstein-Podolsky-Rosen Steering Inequalities with State Information}


\author[one]{James Schneeloch}
\author[one,two]{Curtis J. Broadbent}
\author[one]{John C. Howell}
\address[one]{Department of Physics and Astronomy, University of Rochester, Rochester, NY 14627}
\address[two]{Rochester Theory Center, University of Rochester, Rochester, NY 14627}

\address{}

\begin{abstract}
We discuss the relationship between entropic Einstein-Podolsky-Rosen (EPR)-steering inequalities and their underlying uncertainty relations along with the hypothesis that improved uncertainty relations lead to tighter EPR-steering inequalities. In particular, we discuss how the intrinsic uncertainty in a mixed quantum state is used to improve existing uncertainty relations and how this information affects one's ability to witness EPR-steering. As an example, we consider the recent improvement (using a quantum memory) to the entropic uncertainty relation between pairs of discrete observables (Nat. Phys. \textbf{6}, 659 (2010)) and show that a trivial substitution of the tighter bound in the steering inequality leads to contradictions, due in part to the fact that the improved bound depends explicitly on the state being measured. By considering the assumptions that enter into the development of a steering inequality, we derive correct steering inequalities from these improved uncertainty relations and find that they are identical to ones already developed (Phys. Rev. A, \textbf{87}, 062103 (2013)). In addition, we consider how one \emph{can} use information about the quantum state to improve our ability to witness EPR-steering, and develop a new continuous variable symmetric EPR-steering inequality as a result.
\end{abstract}

\begin{keyword}
EPR steering \sep entanglement \sep EPR-paradox \sep uncertainty relations \sep entropy


\end{keyword}

\end{frontmatter}


\section{Introduction}
\label{Introduction}
Uncertainty relations are used not only to great effect in expressing fundamental limitations of precision measurements; they are also useful in witnessing entanglement through demonstrations of the EPR paradox \cite{EPR1935} by the violation of EPR-steering inequalities\footnote{EPR-steering inequalities are relations illustrating that, if the effect of measurement indeed cannot travel faster than light, then the measurement uncertainties of one party, whether or not they are conditioned on the outcomes of another party, have the same lower bound.}. Since EPR-steering inequalities are derived from underlying uncertainty relations, it is natural to consider whether improved uncertainty relations inevitably lead to improved EPR-steering inequalities. We provide an answer to this question in this letter, as well as explore just how additional information from the state of a system can be used to improve one's ability to witness EPR-steering.

EPR-steering is a form of nonlocality intermediate between Bell-nonlocality and nonseparability \cite{Wiseman2007}. A joint quantum system is said to exhibit EPR-steering (or be EPR-steerable) if its local measurement correlations are sufficiently strong to demonstrate the EPR-paradox \cite{EPR1935}. As a consequence of EPR-steering, consider two parties, Alice and Bob, sharing quantum systems A and B, respectively. Bob can determine that he and Alice share entanglement even when he does not trust Alice's measurements provided A and B are sufficiently entangled. Bob does this by ruling out the possibility that Alice is preparing and sending systems to Bob, and then using her knowledge of those systems to announce fabricated "measurements" she expects to be correlated to Bob's results. In this scenario, the measurement correlations across complementary observables (say, in both position and momentum, or in linear and circular polarizations of light) can only be so high. These models, in which Bob is receiving an unknown, but well-defined quantum state classically correlated to Alice's results are known as models of \emph{local hidden states} (LHS) for Bob. When the measurement correlations across complementary observables is sufficiently high, Bob can rule out all LHS models and verify that he and Alice must be sharing entanglement.

Ruling out LHS models for Bob is done by violating EPR-steering inequalities, i.e., inequalities derived from the necessary form that the joint measurement probabilities must have \eqref{firstoneLHS} in an LHS model (for Bob). Steering inequalities are useful not only because they witness entanglement without needing to perform complete state tomography; they also verify entanglement between two parties even when the measurements of one party are untrusted \cite{Wiseman2007}. For this reason, steering inequalities have been shown to be useful in entanglement-based quantum key distribution \cite{BranciardQKD2012}.

In some cases, improvements to uncertainty relations lead to better EPR-steering inequalities. For example,  Bia{\l}ynicki-Birula and Mycielski's entropic uncertainty relation \cite{BiałynickiBirula1975} for position and momentum encompasses the variance-based Heisenberg uncertainty relation \cite{Heisenberg1927}. Similarly, the resulting entropic EPR-steering inequality \cite{Walborn2011} encompasses the variance-based steering inequality \cite{Reid1989}, permitting EPR-steering to be witnessed in more diverse systems. In spite of this particular example, however, improving uncertainty relations does not necessarily improve steering inequalities, as we shall show.
 
Previously \cite{Schneeloch2013}, we showed how a state-independent entropic uncertainty relation relating a pair of $N$-dimensional discrete observables, say, $\hat{Q}$ and $\hat{R}$, gives rise to a formulation of a corresponding EPR-steering inequality between a pair of systems $A$ and $B$. In particular, given the uncertainty relation
\begin{align} \label{Maassenineq}
&H(Q) + H(R) \geq\log(\Omega),\\ \label{bb}
&:\:\Omega \equiv \min_{i,j} \bigg(\frac{1}{|\langle q_{i}|r_{j}\rangle|^{2}}\bigg),
\end{align}
there is a corresponding EPR-steering inequality,
\begin{equation}\label{discWSE}
H(Q^{B}|Q^{A}) + H(R^{B}|R^{A}) \geq\log(\Omega^{B}),
\end{equation}
where $\Omega^{B}$ is $\Omega$ as defined in equation \eqref{bb}, but for observables on system $B$. Here, $H(Q)$ is the Shannon entropy of the measurement probabilities of observable $\hat{Q}$, i.e.,
\begin{equation}
H(Q)\equiv -\sum_{i}P(q_{i})\log(P(q_{i})),
\end{equation}
where $P(q_{i})\equiv \mathrm{Tr}(\hat{\rho}\; |q_{i}\rangle\langle q_{i}|)$. Similarly, $H(Q^{A},Q^{B})$ is the Shannon entropy of the joint measurement probabilities of observables $\hat{Q}^{A}$ and $\hat{Q}^{B}$, i.e.,
\begin{equation}
H(Q^{A},Q^{B})\equiv -\sum_{i,j}P(q^{A}_{i},q^{B}_{j})\log(P(q^{A}_{i},q^{B}_{j})),
\end{equation}
where $P(q^{A}_{i},q^{B}_{j})\equiv \mathrm{Tr}(\hat{\rho}^{AB}\; |q^{A}_{i}\rangle\langle q^{A}_{i}|\otimes |q^{B}_{j}\rangle\langle q^{B}_{j}|)$. In addition, $H(Q^{B}|Q^{A})$ is the conditional Shannon entropy, where $H(Q^{B}|Q^{A})=H(Q^{A},Q^{B})-H(Q^{A})$, and all logarithms are taken to be base 2. 

Examination of \eqref{Maassenineq} and \eqref{discWSE} suggests that entropic EPR-steering inequalities may be obtained from entropic uncertainty relations by a trivial substitution of conditional entropies for marginal entropies. Indeed, as we shall show below, when the uncertainty bound is state-independent, this strategy is appropriate \footnote{This point was made in a previous publication in which only state-independent uncertainty relations were considered \cite{Schneeloch2013}.}. In contrast, such a substitution is not necessarily appropriate when the uncertainty bound is state-dependent, as we illustrate with the following example.

Recently, Berta \emph{et.~al} \cite{Berta2010} developed an improved entropic uncertainty relation which raises the bound on the right hand side of \eqref{bb} when the von Neumann entropy of Bob's system described by density operator $\hat{\rho}^{B}$ is known,
\begin{equation}\label{improvedrelation}
H(Q^{B}) + H(R^{B}) \geq\log(\Omega^{B}) + S(\hat{\rho}^{B}).
\end{equation}
This improved uncertainty relation is a consequence of Berta \emph{et.~al.}'s uncertainty principle in the presence of quantum memory \cite{Berta2010} \footnote{The improved uncertainty relation \eqref{improvedrelation} has the appealing intuition that if the minimum uncertainty limit when measuring a pure state is given by $\log(\Omega^{B})$, then the minimum uncertainty limit when measuring a mixture of pure states is larger by the intrinsic uncertainty of the mixture.}. With a quantum memory maximally entangled with a system to be measured, the expected outcome of a particular observable of that system can be known with arbitrary precision by measuring the corresponding observable of the entangled memory. The improved uncertainty relation \eqref{improvedrelation} for single systems arises as a special case when the quantum memory is uncorrelated with the system to be measured.

This state-dependent improved uncertainty relation \eqref{improvedrelation} cannot be adapted into an EPR-steering inequality by the substitution of conditional entropies for marginal ones, as doing so would lead to a contradiction. That is,
\begin{equation}
H(Q^{B}|Q^{A})+H(R^{B}|R^{A})\ngeq \log (\Omega^{B}) + S(\hat{\rho}^{B}).
\end{equation}
If $\hat{Q}$ and $\hat{R}$ were mutually unbiased observables, such that $\log(\Omega)=\log(N)$, and the subsystems $A$ and $B$ were maximally mixed, so that $S(\hat{\rho}^{A})=S(\hat{\rho}^{B})=\log(N)$, we would find that the substitution leads to the following inequality,
\begin{equation}\label{contradiction}
H(Q^{B}|Q^{A}) + H(R^{B}|R^{A}) \geq 2 \log(N),
\end{equation}
which is an inequality that separable states can violate. 
As an example, consider the maximally correlated mixed two-qubit state, i.e., the separable state obtained from an even mixture of the separable joint spin-z states $\mid\uparrow\rangle\langle\uparrow\mid\otimes\mid\downarrow\rangle\langle\downarrow\mid$ and $\mid\downarrow\rangle\langle\downarrow\mid\otimes\mid\uparrow\rangle\langle\uparrow\mid$;
\begin{equation}
\hat{\rho}^{AB}=  \begin{pmatrix}
  0 & 0 & 0 & 0 \\
  0 & \frac{1}{2} & 0 & 0 \\
   0 & 0 & \frac{1}{2} & 0 \\
   0 & 0 & 0 & 0 
 \end{pmatrix} :
 \hat{\rho}^{A}= \hat{\rho}^{B}=\begin{pmatrix}
  \frac{1}{2} & 0 \\
  0 & \frac{1}{2} 
 \end{pmatrix}.
\end{equation}
In this system, the alleged inequality \eqref{contradiction} has the form
\begin{equation}\label{sigsig}
H(\sigma_{z}^{B}|\sigma_{z}^{A}) + H(\sigma_{x}^{B}|\sigma_{x}^{A}) \geq 2,
\end{equation}
because $S(\hat{\rho}^{B})=1$ bit. Measurement in the Pauli $\sigma_{z}$ basis, which is the same as the eigenbasis, gives $P(\uparrow,\uparrow)=P(\downarrow,\downarrow)=0$, and $P(\uparrow,\downarrow)=P(\downarrow,\uparrow)=\frac{1}{2}$. Since the measurement result of $\hat{\sigma}_{z}^{A}$ is completely correlated with the result of $\hat{\sigma}_{z}^{B}$, the conditional entropy, $H(\sigma_{z}^{B}|\sigma_{z}^{A})$ is zero bits. Measurement in the Pauli $\sigma_{x}$ basis, which is mutually unbiased with respect to the $\sigma_{z}$ basis, results in a uniform distribution for the joint measurement probabilities, and gives a conditional entropy, $H(\sigma_{x}^{B}|\sigma_{x}^{A})$, of $1$ bit. Since the total on the left hand side of \eqref{sigsig} is one bit less than the bound of 2 bits, we would conclude that this classically correlated separable state is not only entangled, but EPR-steerable. This is a contradiction. To resolve this contradiction, we must examine the assumption of an  LHS model that goes into the derivation of entropic EPR-steering inequalities.

\section{The LHS model assumption with the improved uncertainty bound}
Given a pair of quantum systems $A$ and $B$, we say that the pair admits an LHS model for $B$ if $B$ has a well-defined quantum state only classically correlated with $A$. Such a system can be considered to be ``EPR-local'', and admits the possibility that Alice is preparing and sending systems to Bob and using her knowledge of those systems to announce ``measurements'' correlated to what she believes Bob's outcomes will be. As such, being able to rule out such an LHS model successfully witnesses entanglement between Alice and Bob even when Alice's results are untrusted \cite{Wiseman2007}. 

In \cite{Schneeloch2013}, as well as in \cite{Walborn2011}, the assumption of an LHS model for $B$ is enforced by requiring the joint measurement probabilities to take the following form,
\begin{equation}\label{firstoneLHS}
P(r^{A}_{i},r^{B}_{j}) = \sum_{\lambda} P(\lambda) P(r^{A}_{i}|\lambda)P_{q}(r^{B}_{j}|\lambda) .
\end{equation}
Though this form bears striking resemblance to \emph{local hidden variable} models \cite{CHSHbell1969}, there is the additional assumption that Bob's measurements arise from a quantum probability distribution (denoted by subscript $q$), where $P_{q}(r^{B}_{j}|\lambda)\equiv \mathrm{Tr}\;\big[|r^{B}_{j}\rangle\langle r^{B}_{j}| \hat{\rho}^{B}_{\lambda}\big]$, and is only dependent on the details of the hidden parameter $\lambda$ (governing the possible state prepared by Alice). No such assumption is imposed on Alice's measurements. In this situation, we assume both that Bob's measurements are constrained by quantum uncertainty relations, and that his measurement outcomes are conditionally independent of Alice's results. With these assumptions, we are led to the LHS criterion \cite{Walborn2011,Schneeloch2013}, 
\begin{equation}\label{LHScrit}
H(Q^{B}|Q^{A}) + H(R^{B}|R^{A}) \geq \sum_{\lambda} P(\lambda)\big(H_{q}(Q^{B}|\lambda) + H_{q}(R^{B}|\lambda)\big).
\end{equation}
In \cite{Schneeloch2013}, the derivation of the entropic EPR-steering inequalities is finished by substituting Maassen and Uffink's bound \eqref{bb} into the right hand side of \eqref{LHScrit}, giving us the steering inequality \eqref{discWSE}. 

To develop an improved EPR-steering inequality, with the improved entropic uncertainty relation \eqref{improvedrelation}, we argue that for each value of the hidden variable(s) $\lambda$ governing the preparation of Bob's system, the improved uncertainty relation holds,
\begin{equation}
H(Q^{B}|\lambda) + H(R^{B}|\lambda) \geq\log(\Omega^{B}) + S(\hat{\rho}^{B}_{\lambda}),
\end{equation}
giving us the inequality,
\begin{equation}\label{LHSsum}
H(Q^{B}|Q^{A}) + H(R^{B}|R^{A})\geq\log(\Omega^{B}) + \sum_{\lambda}P(\lambda) S(\hat{\rho}^{B}_{\lambda}),
\end{equation}
for each LHS model given by $\lambda$ and $P(\lambda)$. As it stands, \eqref{LHSsum} is an unsatisfactory steering inequality since the right-hand side retains an explicit dependence on $\lambda$. Instead, we desire an inequality that does not depend on $\lambda$, and which, when violated, rules out all possible LHS models for Bob. In other words, we need to determine a minimal constant $\Delta$, 
\begin{equation}
\Delta\equiv\min_{LHS} \sum_{\lambda}P(\lambda) S(\hat{\rho}^{B}_{\lambda}),
\end{equation}
which, with $\log(\Omega^{B})$, gives the smallest possible sum of conditional entropies that an LHS model for Bob can have,
\begin{equation}\label{LHSsum2}
H(Q^{B}|Q^{A}) + H(R^{B}|R^{A})\geq\log(\Omega^{B}) + \Delta.
\end{equation}
When violated, \eqref{LHSsum2} successfully rules out all LHS models for Bob, demonstrating EPR-steering.

In the following arguments, we show that $\Delta$ must be zero. From there, we see that this particular state-dependent improvement to the uncertainty relation \eqref{improvedrelation} has no effect on the associated steering inequality \eqref{discWSE}. 

Consider that the sum being minimized in $\Delta$ is a weighted sum of von Neumann entropies, $S(\hat{\rho}^{B}_{\lambda})$, of the states Bob is receiving from Alice. These entropies, like all von Neumann entropes must take values between zero and $\log(N)$.  Since Alice is free to send any distribution of states $\hat{\rho}^{B}_{\lambda}$ to Bob, the weighted sum of entropies can also take any value between zero and $\log(N)$, the lower limit being when Alice is sending to Bob a distribution of pure states (with zero von Neumann entropies). Thus in order to have a bound which when violated, rules out all possible LHS models for Bob, $\Delta$ must be zero. Knowing this, $\log(\Omega^{B})$ remains the lower bound for witnessing EPR-steering through \eqref{discWSE}, even when the state $\hat{\rho}^{B}$ can be determined through the use of a quantum memory.

\section{Using state information to improve steering inequalities}
Though the previous state-dependent improved entropic uncertainty relation \eqref{improvedrelation} did not yield an improved EPR-steering inequality, it is straightforward to show that one \emph{can} use information about the state of a quantum system to improve one's ability to witness EPR-steering. To explore this, we note that uncertainty relations can be defined as any physically imposed constraint on measurement probability distributions. Most uncertainty relations are lower bounds on measurement uncertainties, but it is also possible to bound measurement uncertainties from above \cite{SanchezRuiz1995}. As explored in \cite{Schneeloch2013,Schneeloch2012}, upper bounds on measurement uncertainties are used to develop symmetric steering inequalities \footnote{Symmetric steering inequalities are steering inequalities that are symmetric between parties. As such, the violation of a symmetric steering inequality rules out models of local hidden states for both Alice and Bob, allowing both of them to verify entanglement even when neither of them trusts each other's measurements (though they trust their own).} in terms of the mutual information, which is defined for discrete observables as:
\begin{equation}
H(Q^{A}\!:\!Q^{B})\equiv H(Q^{A})+H(Q^{B})-H(Q^{A},Q^{B})
\end{equation}
and for continuous observables as:
\begin{equation}
h(x^{A}\!:\!x^{B})\equiv h(x^{A})+h(x^{B})-h(x^{A},x^{B}).
\end{equation}
Note that for continuous observables, the entropies $h(x^{A})$, $h(x^{B})$, and $h(x^{A},x^{B})$ are differential entropies \cite{Cover2006}, where
\begin{equation}
h(x)\equiv -\int dx\:\rho(x)\log(\rho(x)),
\end{equation}
and $\rho(x)$ is the probability density of continuous random variable $x$.

In \citep{Schneeloch2013}, we used the fact that the  discrete entropy of an $N$-dimensional system is no larger than $\log(N)$ to develop symmetric EPR-steering inequalities using the discrete mutual information. 
From the conditional steering inequality \eqref{discWSE}, we developed the symmetric steering inequality,
\begin{equation}
H(Q^{A}\!:\!Q^{B})+H(R^{A}\!:\!R^{B})\leq\max_{i=\{A,B\}} \log\bigg(\frac{N^{2}}{\Omega^{i}}\bigg).
\end{equation}
For continuous observables, however, there is no known state-independent upper limit to the entropy. We can attempt to derive a symmetric EPR-steering inequality in the same fashion for continuous variables bu subtracting Walborn \emph{et.~al}'s \cite{Walborn2011} conditonal entropic steering inequality,
\begin{equation}\label{walbsteerineq}
h(x^{B}|x^{A})+h(k^{B}|k^{A})\geq \log(\pi e),
\end{equation}
from the sum of marginal entropies $h(x^{B}) + h(k^{B})$, but since this sum of marginal entropies is unbounded for continuous variables, the resulting  sum of mutual informations, $h(x^{A}\!:\!x^{B}) + h(k^{A}\!:\!k^{B})$, is also unbounded. However, by using additional information about the state of the system, we can form an upper bound.

To find an upper bound for the sum of mutual informations, we use the fact that knowledge of measurement statistics allows one to further constrain the measurement probability distributions. In particular, we can bound from above the differential entropy $h(x)$ if we know the variance $\sigma_{x}^{2}$, \cite{BiałynickiBirula1975,Cover2006}, so that
\begin{equation}
h(x)\leq \frac{1}{2}\log(2 \pi e \sigma_{x}^{2}),
\end{equation}
giving us the relation,
\begin{equation}
h(x^{B}) +h(k^{B})\leq \log(2 \pi e \;\sigma_{x^{B}}\sigma_{k^{B}}).
\end{equation}
Subtracting this inequality from \eqref{walbsteerineq}  and symmetrizing gives us the inequality
\begin{equation}\label{mutinginrq}
h(x^{A}\!:\!x^{B})+h(k^{A}\!:\!k^{B})\leq \max_{i=\{A,B\}}\log (2 \sigma_{x^{i}}\sigma_{k^{i}}),
\end{equation}
which we make symmetric by taking the largest bound between Alice and Bob's measurements. 

Incidentally, this leads directly to an experimentally tenable symmetric steering inequality using discrete approximations to the continuous mutual informations similar to the one in \citep{Schneeloch2012} by noting that the mutual information between the discrete approximations of two continuous variables, ($H(X^{A}\!\!:\!\! X^{B})$), is never more than the mutual information between the continuous variables themselves ($h(x^{A}\!:\! x^{B})$) \cite{Schneeloch2013supp}.
\begin{equation}
H(X^{A}\!:\!X^{B})+H(K^{A}\!:\!K^{B})\leq \max_{i=\{A,B\}}\log (2 \sigma_{x^{i}}\sigma_{k^{i}}).
\end{equation}

\section{Conclusion}
We have shown that even substantially improved uncertainty relations do not necessarily lead to improved EPR-steering inequalities when these improvements are state-dependent\footnote{We note that there are state-\emph{independent} improvements to Maassen and Uffink's uncertainty relation when the observables are not fully unbiased, but that Maassen and Uffink's relation remains tight for mutually unbiased observables \cite{deVicente2008,Bosyk2011,Puchala2013,Coles2013}.}. This is due to the fact, that in order to rule out all LHS models, one must find the lowest possible bound given by an LHS model and use that in constructing a valid entropic EPR-steering inequality. However, we have also shown how one \emph{can} use state-dependent information to improve capabilities of witnessing EPR-steering, and have developed the symmetric steering inequality \eqref{mutinginrq} as a result. In addition to finding this lowest possible bound for Berta \emph{et.~al}'s improved uncertainty relation \eqref{improvedrelation}, we've also shown how to find this lowest possible bound in general. This will permit future improvements to uncertainty relations to be easily incorporated into existing EPR-steering inequalities \cite{Schneeloch2013}. Futhermore, we've shown why additional consideration must be taken in developing EPR-steering inequalities from state-dependent uncertainty relations.

We gratefully acknowledge support from DARPA DSO under grant numbers W911NR-10-1-0404 and W31P4Q-12-1-0015. CJB acknowledges support from ARO W911NF-09-1-0385 and NSF PHY-1203931.





\bibliographystyle{elsarticle-num}
\bibliography{EPRbib11}







\end{document}